# PRE-TRAINING FOR LOW RESOURCE SPEECH-TO-INTENT APPLICATIONS

*Pu Wang, Hugo Van hamme*

KU Leuven, Department of Electrical Engineering-ESAT
Kasteelpark Arenberg 10, Bus 2441, B-3001 Leuven Belgium
{pu.wang, hugo.vanhamme}@esat.kuleuven.be

## ABSTRACT

Designing a speech-to-intent (S2I) agent which maps the users' spoken commands to the agents' desired task actions can be challenging due to the diverse grammatical and lexical preference of different users. As a remedy, we discuss a user-taught S2I system in this paper. The user-taught system learns from scratch from the users' spoken input with action demonstration, which ensure it is fully matched to the users' way of formulating intents and their articulation habits. The main issue is the scarce training data due to the user effort involved. Existing state-of-art approaches in this setting are based on non-negative matrix factorization (NMF) and capsule networks. In this paper we combine the encoder of an end-to-end ASR system with the prior NMF/capsule network-based user-taught decoder, and investigate whether pre-training methodology can reduce training data requirements for the NMF and capsule network. Experimental results show the pre-trained ASR-NMF framework significantly outperforms other models, and also, we discuss limitations of pre-training with different types of command-and-control (C&C) applications.

## 1. INTRODUCTION

In this paper we will focus on speech-to-intent (S2I) systems, which are also known as command-and-control (C&C) spoken language understanding (SLU) systems. These systems extract sematic information from spoken input and map them to pre-defined task descriptions. Take the home automation agent as an example, a command "turn on the light of the bedroom" could be mapped to the task: "{*action: switch on, location: bedroom, object: light*}", the agent will therefore conduct the related action which matches the desired task. Such an S2I system could benefit our daily life, especially for physically disabled or elderly people, which help them live more independently.

Conventional approaches to these applications are composed of two separate modules: an automatic speech recognition (ASR) component to get textural transcriptions from input spoken language and a natural language understanding (NLU) component to map intermediate transcripts to the semantics [1]. Such a pipeline structure requires some assumptions about users and their preference since ASR needs to be trained on the specific language or speech conditions. This limitation will introduce errors to the speech-to-intent system when the user speaks an under-resourced language or exhibits a strong dialect. Even stronger problems are experienced by speech-impaired users, especially when the speech impairment is caused by motor disability. Although they could greatly benefit from the speech-to-intent agent, e.g. dysarthric speech inevitably leads to unsatisfactory results of ASR. This in turn leads to degraded input for the NLU component leading to failure of the S2I system [2, 3].

NLU components are typically based on "slot filling", i.e. pre-defined task slots, as aforementioned "{*action, location, object*}", need to be assigned a value. This can be achieved using rule-based systems or with statistical or deep learning approaches [4]. While in the former it is hard to predict the different grammatical and lexical ways a user can express the intent, the latter one requires plenty of utterance-intent pairs to achieve good performances. The main issues come from the diverse nature of speech signals, as one simple task could be expressed by multiple phrases. It is impossible to form a sufficient training set that provides all diversities. Advanced deep learning approaches like RNN may not get convincing performance in the domain of the application.

An alternative approach that is advocated here is the *user-taught direct speech-to-intent system*. "User-taught", refers to a speech-to-intent system that learns to directly map spoken commands to the users' defined tasks from user-provided examples [5-8]. The user speaks a command and then demonstrates the related task through the agents' interface. In our approach, the mapping from speech to task is learned end-to-end, i.e. it gets rid of any intermediate textual representation. The S2I system will hence not degrade by feeding the NLU system with corrupted ASR output that the NLU has not seen during training. Moreover, the system will be matched to the users' way of formulating intents. The disadvantage of user-taught system is that users need to provide training samples. Therefore, in this work, an important evaluation criterion will hence be the amount of training samples required for a given accuracy [7, 8].

In our past work, we have shown that capsule networks achieve state-of-the art performance of this task with limited training data [7, 8]. An older, yet powerful method for this task, resulting in compact models, is based on non-negative matrix factorization (NMF) [5, 6]. These two frameworks mainly deal with the data scarcity issue by designing small-scale models with less-data hungry methods, which is a suboptimal solution.

Pre-training is a successful remedy for the lack of task-specific training data [9-12]. Parts of the system are trained on a different, but related task, for which training data is available in abundance. Other parts of the network are then trained on the task-specific data, possibly followed by a fine-tuning stage of the complete network. Pre-training language representation models such as BERT [9], ALBERT [10], ERNIE framework [11], and XLnet [12] has been proven effective in various language understanding domains including question answering, language inference, sentiment analysis and document ranking [9-12]. Following this idea, we defend a different end-to-end user-taught S2I design which introduces the pre-training model. The efficiency of pre-training will be further explored for our task. A trivial example of a pre-trained S2I system would be to train the ASR component of a S2I system on generic data and train the SLU component on the task/user-specific data.

Therefore, in this paper, we extent the existing NMF/capsule network-based user-taught systems by combining with pre-training methodology. The proposed user-taught system consists of an ASR module which encodes inputs to high character-level posteriorgram features and an intent decoder to decipher the task label from the encoded features. The ASR component is pre-trained by ESPnet [13] on a generic large-scale dataset, and its model parameters are frozen during speech-to-intent mapping. The intent decoders are modifications of prior NMF and capsule network structures which inherit the merit of the original compact model scale. We will discuss the contributions of pre-training regarding these two different intent decoders on two public datasets.

The existing NMF-based and capsule network-based user-taught systems which serve as both the baselines and the designed framework's intent decoders will first be explained in section 2. Section 3 will give the detailed structure of the designed combination framework. In section 4, we will describe the experimental setup for evaluation, and the corresponding results and discussion will be presented in section 5. In section 6 we will conclude our work.

## 2. PRIOR WORK

In this paper, two previous state-of-art speech-to-intent structures will serve as the baselines. The extension with pre-training methods will combine with these two baselines to form a new design.

### 2.1. NMF-based speech-to-intent

The detailed structure of NMF-based speech-to-intent is proposed and extensively discussed in [5, 6]. The main idea of NMF is decomposing a nonnegative data matrix $V$ into two low-rank nonnegative matrixes $W$ and $H$:
$$V \approx WH$$
where each column of $V$ encodes an utterance plus demonstration, $W$ is dictionary whose columns contain recurrent patterns in the data and $H$ is an activation matrix which defines where the recurrent patterns of dictionary occur in the data.

For speech-to-intent application, the model will first learn the dictionary matrix during the training process. Mathematically, the above equation could be specified as:
$$\begin{bmatrix} V_s^{(train)} \\ V_a^{(train)} \end{bmatrix} \approx \begin{bmatrix} W_s \\ W_a \end{bmatrix} H^{(train)}$$
where, the left side of the equation is the utterance-intent training pair which is composed of two parts: The top part $V_s^{(train)}$ is the semantic part which contains a binary many-hot encoding of the intent as demonstrated by the users. Each column of the bottom part $V_a^{(train)}$ encodes the acoustics of a full utterance as a histogram of acoustic co-occurrences (HAC), i.e. the bigram frequency of events at multiple delays [14]. This is a fixed-sized sentence embedding that is sensitive to the order of the acoustic events. $W_s$ and $W_a$ on the right side of the equation are the corresponding semantic and acoustic parts in the dictionary.

During the test process, only the acoustic part of the test sample $V_a^{(test)}$ is available. The activation matrix of the test sample is found by decomposing $V_a^{(test)}$ with the $W_a$ from the training procedure:
$$V_a^{(test)} \approx W_a H^{(test)}$$
The estimated label of the test sample final obtained from:
$$V_s^{(test)} \approx W_s H^{(test)}$$

### 2.2. Capsule network-based speech-to-intent

The detailed structure of capsule network-based encoder-decoder speech-to-intent could be found in [7]. The inputs of the system are first encoded to high level features $F$ by a 2-layer bidirectional GRU.

The encoded features $F$ are then converted to capsule vectors $S_i$ by an attention and distributor mechanism:
$$S_i = Squash(w_s \cdot \sum_t \alpha_t \, \delta_{ti} F_t).$$
Here, $S_i$ is the representation for capsule $i$. *Squash* is soft normalization function in capsule network to ensure the length of $S_i$ lies between 0 and 1. $w_s$ are trainable weights of the squash layer. $\alpha_t$ is the attention weight for each time step, which is used to filter out the unimportant time frames in the sequence (e.g. silence). $\delta_{ti}$ are the distribution weights of distributor to assign each time step $t$ to the hidden capsule $i$.

$\alpha_t$ and $\delta_t$ are calculated from:
$$\alpha_t = sigmoid(w_a \cdot F_t + b_a)$$
$$\delta_t = softmax(w_d \cdot F_t + b_d)$$

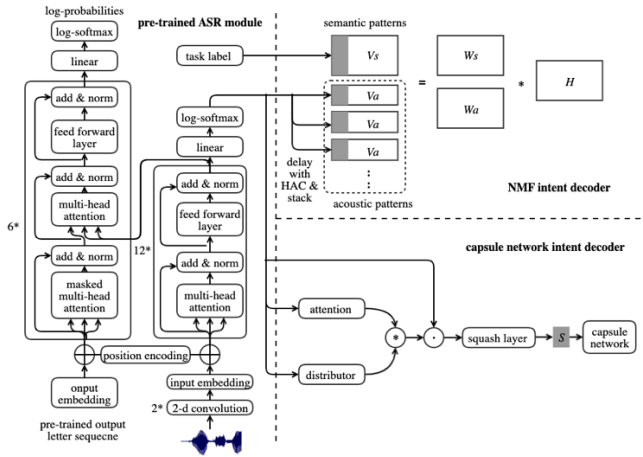

Fugure 1: User-taught speech-to-intent application with pre-training

Here, $w_a$ and $b_a$, $w_d$ and $b_d$ are weights and biases of the sigmoid and softmax layers respectively.

A more extensive description of capsule networks can be found in [15]. In general, a capsule is characterized by a vector with length between 0 and 1 which represents the probability of the capsule's (presented label) occurrence, and orientation contains the latent information of the capsule.

## 3. MODEL

The overall structure of our combination model is shown in Figure 1. There are two major components: the pre-trained ASR encoder component which serves as the speech feature encoder and the intent decoder. The intent decoder are modifications of NMF and capsule network structure from [6] and [7].

### 3.1. Pre-trained ASR module

The end-to-end character-based ASR module is pre-trained by ESPnet on a large speech corpus following the hybrid ctc/attention procedure of [13, 16-18]. The acoustic model takes 80-dimensional MEL filterbanks with 3 pitch-related features as inputs. To the inputs, positional encodings are added to provide timing information. A 2-layer 2-dimensional convolutional layer with (3,3) kernel size and 256 channels is introduced to do 4-fold subsampling in time. The encoder and decoder are 12-layer and 6-layer transformers respectively with 4 heads of 125-dimensional keys, queries and values. There are 2048 units within each position-wise feedforward layer. Drop out with rate 0.1 is applied to each output layer.

We take outputs of the encoder followed by one linear mapping to get the probabilities of characters. The intent decoder takes the log posterior grams output of the acoustic model as its input.

### 3.2. Intent decoder

We employ two existing decoders: NMF and capsule network to investigate whether pre-training would benefit both models.

For the NMF decoder, the probability of the acoustic events (a character) are directly obtained from the encoder. Furthermore, HAC is utilized with delays 1, 2, 3, 5 to encode timing information.

For capsule network decoder, the encoded log-features will direct input to attention and distributor mechanism followed by 2-layer capsule network. There are 32 hidden capsules with 64 dimensions in the primary capsule layer and one output capsule for each output label with 8 dimensions in the output capsule layer. Since the original 2-layer GRU module for feature encoded is discarded, the decoder based on capsule network do not contain timing information.

## 4. EXPERIMENT

### 4.1 Validation dataset

The user-taught systems will be tested on two public datasets.

Grabo [19] records 6000 commands to robots from 10 Dutch speakers and 1 English speaker, which contains 33 tasks including positions, movement speeds and actions, like "drive quickly to the right". Therefore, the order of each word does not really matter, since they will not strongly change the meaning even with disordered organization.

PATience CORpus (Patcor) [20, 21] records 2000 commands for a vocally guided card game from 8 Dutch speakers, which contains 38 tasks related to the value and the suit of the card being moved or the target positions, like "put card $x$ on card $y$". Therefore, the word order information of this dataset is considerable relevant.

### 4.2 Pre-trained dataset

The S2I evaluation data are in Dutch, so the end-to-end ASR encoder is pre-trained on the Flemish part of Corpus Gesproken Nederlands (CGN) [22]. We included all data except that narrow-band recordings (components c and d) and the spontaneous conversations (components a). With data augmentation with additive noise and reverberation, there is about 700 hours of training data.

### 4.3 Experimental setup

The NMF-based and capsule network-based speech-to-intent application in [6] and [7] will serve as the basslines, the hyper parameters of these two structures are chosen as in [6, 7] without alteration.

We mainly compare the learning curves of the accuracy for intent label classification under the speaker-dependent setting. The learning curve records the model's performance on an increasing amount of training data and tests on all remaining data using 5-fold cross-validation for each speaker. The utterances from each speaker are randomly shuffled and are divided into 15 blocks and 5 blocks for Grabo and Patcor respectively [7]. The final learning curve is the average accuracy across all speakers.

## 5. RESULTS

In Figure 2, the average accuracy results of different models are plotted as a function of the amount of training samples. For the Grabo data, both decoders combined with the pre-training method outperform the baselines without pre-training, Moreover, the pre-trained ASR-NMF framework significantly improves the performance in general. For the Patcor data, the NMF decoder with pre-trained ASR is also the best, but capsule network does not benefit from pre-training.

Such differences could be explained by the pre-trained encoded features containing inadequate or insufficient information about timing for the capsule network. As aforementioned, NMF encodes time through the HAC, while the capsule network is immediately taking weighted averages of the inputs features. Despite the fact that the pre-trained encoder uses positional encoding, it may not result in features that have an adequate representation for time, which the capsule network can use.

On the Patcor dataset, where word order is relevant to the semantics, the capsule network loss in accuracy due to insufficient timing information is not compensated by the benefits from pre-training and the accuracy drops considerably. For the Grabo dataset, task label for commands like "drive quickly to the right" stay the same with different word order like "quickly drive right". Therefore, its performance is less sensitive to the timing information. Here we observe only a minor influence on performance.

This theory is further verified by introducing different amount of order information in the NMF models on the two datasets. Since the NMF decoder includes order information by stacking different HAC delays in the acoustic representations, more stacked delays contain longer timing information. Therefore, we compared accuracy results of pre-training-NMF model with HAC delays 1 to delays 1, 2, 3, 5 which is from 40ms up to 200ms. The results are shown in Figure 3.

From Figure 3, it is clear that models exploiting more timing information (delays 1, 2, 3, 5) work best on both datasets. However, the performance lift by introducing more order information of the Grabo data is very limited (only up to 0.2% improvement), while the performance gain of the Patcor data is significant (up to 5%) which matches our hypothesis: the Patcor data is more sensitive to order than the Grabo data.

## 6. CONCLUSION

In this paper we extended prior NMF/capsule network based-SLU by introducing pre-training methodology to form a new combination framework for low resource user-taught speech-to-intent application. The proposed S2I framework consists of a pre-trained ASR module for features encoding and the NMF/capsule network based intent decoder for features-to-intent mapping. We investigate the contributions of pre-training to NMF and capsule network on two types of commend-to-control data sets: Grabo and Patcor, where the former is insensitive to sequence order and in the latter order information is highly relevant.

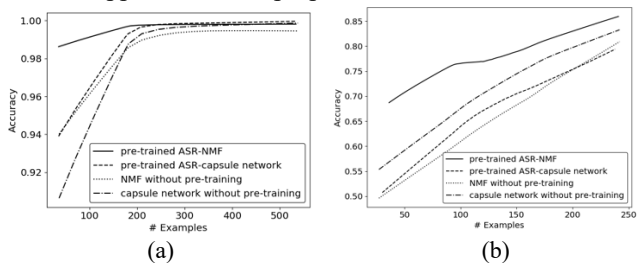

Figure 2: The accuracy curve of compared user-taught systems for (a) the Grabo dataset; (b) the Patcor dataset.

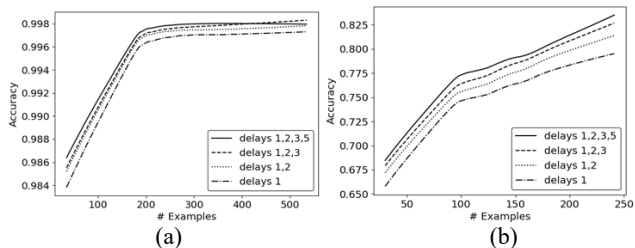

Figure 3: The accuracy curve of different HAC delays in pre-trained ASR-NMF model for (a) the Grabo dataset; (b) the Patcor dataset.

From the experimental results, we can conclude that NMF highly benefits from pre-training: the pre-trained ASR-NMF achieves significant performance gains on both datasets. But pre-trained features are not suitable for the capsule network design by the lack of essential timing information exploitable by capsules. In future work, we will look closely at approaches that can lift such timing information limitations of the pre-trained features.

## 7. ACKNOWLEDGES

The research was supported by the program of China Scholarship Council No. 201906090275, KUL grant CELSA/18/027 and the Flemish Government under "Onderzoeksprogramma AI Vlaanderen".

.